\input harvmac.tex
\noblackbox
\newcount\figno
\figno=0
\def\fig#1#2#3{
\par\begingroup\parindent=0pt\leftskip=1cm\rightskip=1cm\parindent=0pt
\baselineskip=11pt
w^{\prime}lobal\advance\figno by 1
\midinsert
\epsfxsize=#3
\centerline{\epsfbox{#2}}
\vskip 12pt
{\bf Fig. \the\figno:} #1\par
\endinsert\endgroup\par
}
\def\figlabel#1{\xdef#1{\the\figno}}
\def\encadremath#1{\vbox{\hrule\hbox{\vrule\kern8pt\vbox{\kern8pt
\hbox{$\displaystyle #1$}\kern8pt}
\kern8pt\vrule}\hrule}}
\overfullrule=0pt
%
\def\tilde{\widetilde}

\def\inbar{\,\vrule height1.5ex width.4pt depth0pt}
\def\IB{\relax{\rm I\kern-.18em B}}
\def\IC{\relax\hbox{$\inbar\kern-.3em{\rm C}$}}
\def\ID{\relax{\rm I\kern-.18em D}}
\def\IE{\relax{\rm I\kern-.18em E}}
\def\IF{\relax{\rm I\kern-.18em F}}
\def\IG{\relax\hbox{$\inbar\kern-.3em{\rm G}$}}
\def\IH{\relax{\rm I\kern-.18em H}}
\def\II{\relax{\rm I\kern-.18em I}}
\def\IK{\relax{\rm I\kern-.18em K}}
\def\IL{\relax{\rm I\kern-.18em L}}
\def\IM{\relax{\rm I\kern-.18em M}}
\def\IN{\relax{\rm I\kern-.18em N}}
\def\IO{\relax\hbox{$\inbar\kern-.3em{\rm O}$}}
\def\IP{\relax{\rm I\kern-.18em P}}
\def\IQ{\relax\hbox{$\inbar\kern-.3em{\rm Q}$}}
\def\IR{\relax{\rm I\kern-.18em R}}
\def\IT{\relax T}
\font\cmss=cmss10 \font\cmsss=cmss10 at 7pt
\def\IZ{\relax\ifmmode\mathchoice
{\hbox{\cmss Z\kern-.4em Z}}{\hbox{\cmss Z\kern-.4em Z}}
{\lower.9pt\hbox{\cmsss Z\kern-.4em Z}}
{\lower1.2pt\hbox{\cmsss Z\kern-.4em Z}}\else{\cmss Z\kern-.4em Z}\fi}
\def\IGa{\relax\hbox{${\rm I}\kern-.18emw^{\prime}$}}
\def\IPi{\relax\hbox{${\rm I}\kern-.18em\Pi$}}
\def\ITh{\relax\hbox{$\inbar\kern-.3em\Theta$}}
\def\IOm{\relax\hbox{$\inbar\kern-3.00pt\Omega$}}

\font\zfont = cmss10 

\def\bigone{\hbox{1\kern -.23em {\rm l}}}
\def\ZZ{\hbox{\zfont Z\kern-.4emZ}}

\def\CV{{\cal V}}

\def\CN{{\cal N}}

\def\IR{\relax{\rm I\kern-.18em R}}
\def\Dsl{\,\raise.15ex\hbox{/}\mkern-13.5mu D} 
\def\Gsl{\,\raise.15ex\hbox{/}\mkern-13.5mu G} 
\def\Csl{\,\raise.15ex\hbox{/}\mkern-13.5mu C} 

\font\cmss=cmss10 \font\cmsss=cmss10 at 7pt
%
%
\lref\bottcatt{R. Bott and A.S. Cattaneo, ``Integral
invariants of 3-manifolds,'' dg-ga/9710001.}
\lref\wittfive{E. Witten, ``Five-Brane Effective Action In M-Theory,''
hep-th/9610234.}
\lref\freed{D. Freed, ``Anomalies, $p$-forms, and $M$ theory,'' in
preparation.}
\lref\cnhvy{C.G. Callan and J.A. Harvey, ``Anomalies and  Fermion  Zero Modes
on Strings and Domain Walls,'' Nucl. Phys. {\bf B250} (1985) 427.}
\lref\orlando{O. Alvarez, ``Topological Quantization and Cohomology,'' Commun.
Math. Phys. 100 (1985) 279.}
\lref\bt{R. Bott and L.W. Tu, ``Differential Forms in Algebraic Topology,''
Springer-Verlag, 1982, New York.}
\lref\dlm{M.J. Duff, J.T. Liu and R. Minasian, ``Eleven dimensional
origin of string/string duality: a one loop test,'' Nucl. Phys. {\bf B452}
(1995) 261, hep-th/9506126.}
\lref\dscnti{M.F. Atiyah and I.M. Singer,
``Dirac operators coupled to vector bundles,''
Proc. Natl. Acad. Sci. {\bf 81} (1984) 2597.}
\lref\dscntii{
 L. Faddeev and S. Shatashvili,  ``Algebraic and Hamiltonian Methods in the
theory of Nonabelian Anomalies,'' Theor. Math. Fiz., {\bf 60 }
(1984) 206; english transl. Theor. Math. Phys.
{\bf 60} (1984) 770.}
\lref\dscntiii{B. Zumino,
``Chiral anomalies and differential geometry,''
in {\it Relativity, Groups and Topology II},
proceedings of the
Les Houches summer school, B.S. DeWitt and R. Stora, eds.
North-Holland, 1984.}
\lref\dscntiv{
For reviews see {\it Symposium on Anomalies, Geometry and Topology }
W.A. Bardeen and A.R. White, eds. World Scientific, 1985, and
L. Alvarez-Gaum\'e and P. Ginsparg,
``The structure of gauge and gravitational anomalies,''
Ann. Phys. {\bf 161} (1985) 423.}
\lref\msw{J. Maldacena, A. Strominger and E. Witten, ``Black
Hole  Entropy in M theory,'' hep-th/9711053.}
\lref\henningson{M. Henningson, ``Global anomalies in M-theory,''
hep-th/9710126.}
\lref\bonora{ L. Bonora,  C.S. Chu,  M. Rinaldi,
``Perturbative Anomalies of the M-5-brane,''  hep-th/9710063;
``Anomalies and Locality in Field Theories and M theory,''
hep-th/9712205.}
\lref\dealwis{S.P. de Alwis,
``Coupling of branes and normalization of effective actions in
string/M-theory,'' hep-th/9705139.}
\lref\ght{G.W. Gibbons, G.T. Horowitz, and P.K. Townsend,
``Higher-dimensional resolution of dilatonic black-hole singularities,''
hep-th/9410073; Class. Quant. Grav.12 (1995) 297.}
\lref\horowitz{G.T. Horowitz  and S.F. Ross,
``Possible Resolution of Black Hole Singularities from Large N Gauge Theory,''
hep-th/980308.}
\lref\lhl{S. Cordes,  G. Moore,  S. Ramgoolam,
``Lectures on 2D Yang-Mills Theory, Equivariant Cohomology and Topological
Field Theories,''
hep-th/9411210.}
\lref\fkm{S. Ferrara, R.R. Khuri and R. Minasian
``M Theory on a Calabi-Yau Manifold,''  Phys.Lett. { \bf B375} (1996) 81.}
%
%
%

\Title{\vbox{\baselineskip12pt
\hbox{YCTP-P5-98 }
\hbox{EFI-98-10}
\hbox{hep-th/9803205}
}}
{\vbox{\centerline{
Gravitational Anomaly Cancellation}
\centerline{ for  }
\centerline{ $M$-Theory Fivebranes
}}}

\centerline{Dan Freed\foot{School of Mathematics, Institute for
Advanced Study, Olden Lane, Princeton, NJ 08540, permanent address:
Dept. of Mathematics, University of Texas, Austin, TX 78712},~ Jeffrey A.
Harvey\foot{Enrico Fermi
Institute, 5640 Ellis Avenue, University of
Chicago, Chicago, IL 60637},~ Ruben Minasian$^{3}$ and
Gregory Moore\foot{Department of Physics, Yale University,
New Haven, CT 06520}}

\bigskip
\centerline{\bf Abstract}

We study gravitational anomalies for fivebranes in M theory. We show that
an apparent anomaly in diffeomorphisms acting on the normal
bundle is cancelled by a careful treatment of the M theory
Chern-Simons coupling in the presence of fivebranes. One interesting
aspect of our treatment is the way in which a magnetic object
(the fivebrane) is smoothed out through coupling to gravity and
the resulting relation between antisymmetric tensor gauge transformations
and diffeomorphisms in the presence of a fivebrane.

\Date{March 25, 1998}

%
\newsec{Introduction}

M theory is believed to be a consistent theory of quantum gravity
which at low energies reduces to the unique supergravity theory
in $D=11$ spacetime dimensions. M theory contains two types of
extended objects, membranes and fivebranes. Membranes have
odd-dimensional worldvolumes and so there are no anomalies
associated with the zero modes of a membrane. Fivebranes
on the other hand have even dimensional worldvolumes and
chiral zero modes so a computation is needed to see if
there are anomalies in the presence of fivebranes.

A fivebrane of M theory with worldvolume $W_6$ embedded into
eleven-dimensional spacetime $M_{11}$ breaks the Lorentz symmetry
from $SO(10,1)$ to $SO(5,1) \times SO(5)$.
\foot{Since we will be considering
fermions we should really be discussing the covering groups
$Spin(n)$, this distinction will not be important in what follows}
If we believe that
M theory is a well defined theory then diffeomorphisms or equivalently
local Lorentz transformations which map the fivebrane to itself should
be symmetries of the theory.

Diffeomorphisms preserving the fivebrane worldvolume  $W_6 \rightarrow W_6$ are
generated by vector fields
acting either as diffeomorphisms
of the fivebrane worldvolume $W_6$ or as $SO(5)$ gauge transformations
on the connection on the  normal bundle. Using the
metric the normal bundle may be regarded as
a bundle with metric and connection and
structure group $SO(5)$.
The potential anomalies
in worldvolume diffeomorphisms and $SO(5)$
gauge transformations have two obvious sources.
The first is the presence
of chiral zero modes on the fivebrane worldvolume. For a charge one
fivebrane the zero modes consist of a tensor multiplet of $(2,0)$,
$D=6$ supersymmetry. The chiral fields in this multiplet consist
of a chiral fermion transforming in the spinor representation of
$SO(5)$ and a two-form potential with anti-self-dual field strength
which is a singlet under $SO(5)$. The anomaly due to these zero modes
can be computed from the standard descent
formalism
\refs{\dscnti,\dscntii,\dscntiii,\dscntiv}
 and is determined by descent on an
eight-form $I_8^{zm}$. That is, we have $I_8 = d I_7^{(0)}$ and
$\delta I_7^{(0)} = d I_6^{(1)}$ and the anomaly is given by
\eqn\two{
2 \pi \int_{W_6} {I_6^{zm}}^{(1)}. }

The second source of anomalies comes from the presence in supergravity
of a coupling
\eqn\three{\Delta S = \int_{M_{11}} C_3 \wedge I_8^b(R)}
with $I_8^b$ a specific eight-form constructed out of the curvature
on $M_{11}$. Integrating by parts and taking the variation of this
term gives
\eqn\four{ \delta \Delta S = \int_{M_{11}} d G_4 \wedge {I_6^b}^{(1)} .}
The fivebrane of M theory acts as a magnetic source for the three-form
potential $C_3$ of M theory. With $G_4 = d C_3 $ the corresponding
field strength this means, roughly speaking, that
\eqn\five{d G_4 = 2 \pi \delta_5}
where $\delta_5$ is a five-form which integrates to one
in the directions transverse to the fivebrane and has delta
function support on the fivebrane\foot{We will soon give a much more
precise definition of $\delta_5$}.  We thus have for the total
gravitational anomaly
\eqn\six{
2 \pi \int_{W_6} \left( I_6^{zm} + I_6^b \right)^{(1)} =
2 \pi
\int_{W_6} \left( {p_2(N) \over 24} \right)^{(1)} }
with $p_2(N)$ the second Pontrjagin class of the normal bundle \wittfive.
The fact that anomalies in diffeomorphisms of the tangent bundle
cancel between these two sources was pointed out in \dlm.
If we believe that M theory exists and that the
fivebrane of M theory is a well defined
object then there must be some additional mechanism
which cancels the anomaly in diffeomorphisms of the normal bundle. The
cancellation of the normal bundle anomaly
has been investigated in \refs{\dealwis, \bonora, \henningson}, but
a completely satisfactory answer has not yet emerged.

In field theory there are many examples where a smooth soliton solution
of the field theory has chiral zero modes with an anomaly which is
cancelled by inflow from the bulk \cnhvy. This cancellation is
inevitable in field theory since if the original theory was consistent
then the effective action must make sense for any background fields
including those of a smooth soliton. In theories including gravity
the situation is more problematic. The extremal fivebrane of M
theory is non-singular, but becomes singular when perturbed
\refs{\ght,\horowitz}.
In order to study anomalies it is necessary to study not just
a particular fivebrane configuration but families of fivebrane
configurations.

There are two related ways to understand the need for families of
fivebranes. First, the problem of anomalies is the problem of
defining the effective action $e^{i S_{eff}}$ as a function of
the fields of the theory.
Thus the effective action is a section of a line bundle over field
space. The anomaly vanishes if
we can trivialize this bundle. Studying
this question involves studying families of field configurations.
{}From a local point of view anomalies involve a  lack of conservation
of a current or of the energy-momentum tensor in the presence of
background fields. In order to study this conservation we have to
turn on gravitational fields in addition to the background fields
of the fivebrane. Since the fivebrane becomes singular
when we vary the metric it is far from clear that the anomalies
must in fact vanish. A direct approach would involve evaluation of
the Rarita-Schwinger operator in backgrounds with a horizon and
singularity and would be problematic if not impossible.

In this paper we will not try to study the fivebrane directly
as a solution of $D=11$ supergravity. Rather we will
study the fivebrane
as a magnetic source for the three-form potential of M theory
and will divide the fields up into bulk fields and zero mode
fields which are localized on the fivebrane. We will show
that a careful treatment in this framework allows us to
understand the cancellation of all anomalies. We leave
to the future the very interesting question of the relation
of this approach to that based on a direct study of solutions
to supergravity.

We conclude this introduction with a brief comment about the descent
formalism.  In the physics literature on anomalies one commonly writes
Chern-Weil forms such as $I_8$ above as differentials of Chern-Simons forms
$I^{(0)}_7$.  This is valid globally if we fix a reference trivial
connection, but in general the Chern-Simons forms only exist locally.
In the
supergravity theory discussed here we do not want to impose unnecessary
global restrictions on the spacetime and the fivebrane, so the descent
equations are only valid locally.  We make some brief comments about the
global structure in section 4.  A more complete treatment, together with an
exposition of the anomaly cancellation in that global framework, will appear
in~\freed.

In this paper we will follow the conventions and normalizations
of \wittfive. In order to suppress many factors of $2 \pi$ in
various formulae we define $ \Csl_3 = C_3/2 \pi$ and $\Gsl_4 = G_4/2 \pi$
with $C_3$ the three form potential of M theory and $G_4$ its
field strength.

\newsec{The fivebrane source}

Consider a fivebrane of M theory located at $y^{a}=0$, $a=1,2, \ldots 5$
and with longitudinal coordinates $x^{\mu}$, $\mu = 0,1, \ldots 5$. The
most naive expression for the Bianchi identity in the presence of the
fivebrane is
\eqn\seven{d \Gsl_4 = \delta(y^1) \cdots \delta(y^5) dy^1 \wedge \cdots
\wedge d y^5 .}
The quantity on the right hand side is a five-form with integral one
over the transverse space and delta function support on the fivebrane.
However as discussed above, we need to consider families of metrics
so the above expression could at most be correct locally. While a delta
function source is sufficient for computations where $C_3$ enters
linearly, as in \three, we will soon encounter a Chern-Simons term
which is cubic in $C_3$. In order to
have a completely well defined  and non-singular
prescription in such cases we need to smooth out the delta function source.
Having done this we will see that
in the presence of a non-zero $SO(5)$ connection on the normal bundle
we will have to modify  the right hand side
 of \seven\ in order that it
transform covariantly under $SO(5)$ gauge
transformations.

In order to define the fivebrane more carefully
we first use the metric to define a radial
direction away from the fivebrane and we cut out a disc of radius
$\epsilon$ around the fivebrane. That is, we remove a tubular neighborhood of
the fivebrane of radius $\epsilon$. Let $D_\epsilon(W_6)$ denote the total
space
of the resulting disc bundle with base $W_6$ and fibers the discs of radius
$\epsilon$. We will define all bulk integrals as limits as $\epsilon$ goes
to zero of integrals over $M_{11} - D_\epsilon(W_6)$:
\eqn\eight{\int_{M_{11}} {\cal L} \equiv
\lim_{\epsilon \rightarrow 0} \int_{M_{11}-D_\epsilon(W_6)} {\cal L} .}
We will later integrate by parts and use the fact the
$10$-dimensional boundary of
$M_{11} - D_{\epsilon}(W_6)$ is the total space of the $S^4$- sphere bundle
over $W_6$ of radius $\epsilon$,
whose total space we denote $S_{\epsilon}(W_6)$.

In order to smooth out the fivebrane source we choose a smooth function of the
radial direction with transverse
compact support near the fivebrane, $\rho(r)$,  with
$\rho(r) = -1$ for sufficiently small $r$ and
$\rho(r) =0$ for sufficiently large $r$.
The bump form $d \rho$ then has integral  one in the
radial direction.
The smoothed form of \seven\ should then read
\eqn\nine{ d \Gsl_4 = d \rho \wedge e_4/2, }
where $d e_4=0$, $e_4$ is gauge invariant under $SO(5)$ transformations
of the normal bundle, $e_4/2$ has integral one over the fibers of
$S_{\epsilon}$, and $d \rho \wedge e_4/2$ should reduce to the naive expression
on the r.h.s of \seven\ for a flat infinite fivebrane when $d \rho$
approaches a delta function. Physically what we are doing
is smoothing out the magnetic charge of the fivebrane to a
sphere of magnetic charge linking the horizon.

The construction of the smoothed out source involves
standard  mathematics \bt. The
right hand side of \nine\ involves differential forms which arise in a
geometric construction of the Thom class of an oriented vector bundle, in
this case the normal bundle to $W_6$ in $M_{11}$.
As described in \bt\ we may identify the total space of
the normal bundle with a tubular neighborhood
of $W_6$ in $M_{11}$. With this identification the
differential form
$ d \rho \wedge e_4/2$ represents
the Thom class of the
normal bundle and $e_4/2$ is the global angular form. Although the
properties of $e_4$ follow from general principles, an explicit local
formula is useful for constructing explicit objects which will
later appear in the M theory action.

We have $E \rightarrow W_6$ a rank 5 real vector bundle with
metric and connection.
Let $P \rightarrow W_6$ be the principal $SO(5)$
bundle associated to the rank 5 bundle $E$.
Following \bottcatt\ we work on $P \times S^{4}$
and construct a basic form which descends to the sphere bundle
$S(E)$. Think of $S^{4}\subset \IR^{5}$,
choose coordinates $y^a$ for $\IR^{5}$ and
let $\hat y^a \equiv y^a/r$. Of
course, $\hat{y}^a$ is defined only outside of $0\in \IR^{5}$, which
corresponds to the complement of the zero section of~$E$.
On that complement
the pullback of $E$ has a tautological line subbundle, and a perpendicular
oriented 4-plane bundle which we call $F$. Readers with less tolerance
for mathematics can pick a gauge and with little harm done simply
think of the $\hat y^a$ as isotropic coordinates on the $S^4$ fibres
of $S(E)$.

The $SO(5)$ bundle is equipped with a
globally defined connection $\Theta^{ab}= - \Theta^{ba}$.
(We identify ${\bf so(5)}\cong \Lambda^2 \IR^{5}$. )
The Lie algebra ${\bf so(5)}$ acts on
$P \times S^{4}$ in the standard way and
we have horizontal forms:
\eqn\horizontal{
\eqalign{
(D \hat y)^a & \equiv d\hat y^a - \Theta^{ab} \hat y^b\cr
F^{ab} & = d \Theta^{ab} - \Theta^{ac} \wedge \Theta^{cb}. \cr}
}

We now consider the forms:
\eqn\volumi{
\epsilon_{a_1 \cdots a_{5} } (D\hat y)^{a_1} \cdots
(D \hat y)^{a_{4}} \hat y^{a_{5}}
}
\eqn\volumii{
\epsilon_{a_1 \cdots a_{5} } F^{a_1 a_2}\wedge
(D\hat y)^{a_3}
(D \hat y)^{a_{4}} \hat y^{a_{5}}
}
and
\eqn\volumiii{
\epsilon_{a_1 \cdots a_{5} } F^{a_1 a_2}\wedge
F^{a_3 a_4 }  \hat y^{a_{5}}.
}
These forms are all annihilated by
$\iota(X), \CL(X)$, for $X\in {\bf so(5) }$, where
$\iota(X)$ is the contraction and $\CL(X)$ is
the Lie derivative with respect to the vector
field $X$. It follows that these forms
  are basic and descend to $S(E)$. Moreover,
\volumi\ restricts to the volume form on the
$S^{4}$ fiber and thus reduces to the naive expression
in \seven\ in the appropriate limit. However, \volumi\ is not closed.
One can use the identities
\eqn\bianchi{
\eqalign{
d \hat y^a & = \Theta^{ab} \hat y^b + (D\hat y)^{a } \cr
d (D\hat y)^{a } & = \Theta^{ab} (D\hat y)^{b} - F^{ab} \hat y^b\cr
d F^{ab} & = - F^{ac} \Theta^{cb} + F^{bc} \Theta^{ca} \cr}
}
plus the rotational invariance of the forms and
the fact that $\hat y^a (D\hat y)^{a} = 0 $ to show
that up to an overall scale
there
is a unique closed linear combination of \volumi\ - \volumiii.

Equivalently, the curvature of the oriented 4-plane bundle $F$
defined above is the restriction to $F$ of the curvature of $E$ minus a
second fundamental form term.
The Pfaffian of this curvature is represented
by the basic 4-form
\eqn\explctefour{
\eqalign{
e_4(\Theta)   = {1 \over  64 \pi^2}
 \Biggl( \epsilon_{a_1 \cdots a_{5} }
&
(D\hat y)^{a_1} (D\hat y)^{a_2}(D\hat y)^{a_3}
(D \hat y)^{a_{4}} \hat y^{a_{5}} \cr
- 2 \epsilon_{a_1 \cdots a_{5} } F^{a_1 a_2}\wedge
(D\hat y)^{a_3}  \wedge
(D \hat y)^{a_{4}} \hat y^{a_{5}}
+ &
\epsilon_{a_1 \cdots a_{5} } F^{a_1 a_2}\wedge
F^{a_3 a_4 }
  \hat y^{a_{5}} \Biggr).\cr}
}
One can
apply the standard descent formalism to
expressions of the form \explctefour. For example,
assuming that the normal bundle is trivial and choosing
$\Theta=0$ as a basepoint reference connection we have
\eqn\dminone{\eqalign{
e_3^{(0)}(\Theta, \hat y)   = & {1 \over 32 \pi^2}  \epsilon_{a_1 \cdots a_{5}
}
\Biggl(\Theta^{a_1 a_2}  d\Theta^{a_3 a_4}  \hat y^{a_5}  \cr
& \,\, -  {1 \over 2} \Theta^{a_1 a_2} \Theta^{a_3 a_4 } d\hat y^{a_5} - 2
\Theta^{a_1 a_2}  d\hat y^{a_3}  d\hat y^{a_4} \hat y^{a_5} \Biggr).  \cr}}
More generally one can write such formulae for the
difference of two Chern-Simons forms for two connections
$\Theta_1, \Theta_2$ on $E$.

The gauge transformations $\delta \Theta^{a_1 a_2} = (D  \varepsilon)^{a_1
a_2}$ and $\delta \hat y^a = \varepsilon^{a a^{\prime}} \hat y^{a^{\prime}}$
give
\eqn\deterb{e_2^{(1)}(\varepsilon, \Theta, \hat y)  =  {1 \over 16 \pi^2}
\epsilon_{a_1 \cdots a_{5} }
\Biggl(\varepsilon^{a_1 a_2}  d\hat y^{a_3}  d\hat y^{a_4} \hat y^{a_5}  -
\varepsilon^{a_1 a_2}  \Theta^{a_3 a_4}  d\hat y^{a_5}  \Biggr).}
The above expressions have natural generalizations to
all real oriented
bundles of odd rank. We give the general formulae
in the appendix.

While we do not see a direct connection between the analysis
presented here and the discussion in  \wittfive\ concerning
the normal bundle anomaly, it is interesting to
note that the last term in
\explctefour\ is very close to the expressions
which appear in the discussion there.
We expect that upon dimensional reduction our mechanism becomes equivalent to
the
anomaly cancellation mechanism for the IIA fivebrane described in
\wittfive, but we have not worked out the details of this.

\newsec{Connection to Anomalies}

In giving a precise definition of the fivebrane source we encountered
the global angular form $e_4/2$. It is clear from \explctefour\ that
the  global angular form depends on the
connection on the normal bundle and is closed and gauge invariant
under $SO(5)$ gauge transformations acting on the normal bundle. As
described above we
can thus apply descent:
\eqn\zza{e_4 = d e_3^{(0)}, \qquad \delta e_3^{(0)} = d e_2^{(1)} .}

We will now express the uncanceled anomaly \six\ in terms 
of \zza\ using a
result of Bott and Cattaneo \bottcatt.  
Consider a real vector bundle
$N\rightarrow M$ of odd rank $2n+1$, and for convenience 
fix a metric. 
Let $\pi : S(N) \rightarrow M$ denote the unit sphere bundle in $N$.  
Then the lift
$\pi^*N$ has a tautological line subbundle~$L$, and in 
\bottcatt\ it is shown
that the Euler class of the orthogonal complement 
$L^\perp$ satisfies 
$\pi_*
[e_{2n}(L^\perp)^3] = 2 p_{n}(N)$. 
The factor of $2$ is the Euler
characteristic of the even dimensional sphere.  
At the level of cohomology,
this formula follows from a simple argument using the 
splitting principle.
If $N$ has an orthogonal connection, then we represent 
real characteristic
classes as differential forms using Chern-Weil representatives.  
The formula
also holds at the level of differential forms, since 
both sides are gauge
invariant and depend on only a finite number of 
derivatives of the
connection. Applying the result of \bottcatt\ to our 
case, and applying the
descent formalism we have:
\eqn\zzb{{1 \over 6} \int_{S_{\epsilon}(W_6)} {e_4 \over 2} \wedge
{e_4 \over 2} \wedge {e_2^{(1)} \over 2} = \int_{W_6} {(p_2(N))^{(1)} \over 24
}. }

The $D=11$ supergravity which describes the low-energy limit of M
theory contains a Chern-Simons term
\eqn\zzc{S_{CS} =  -{2 \pi  \over 6} \int_{M_{11}} \Csl_3 \wedge d \Csl_3
\wedge d \Csl_3 =
{2\pi \over 6} \int_{M_{12}} d \Csl_3 \wedge d \Csl_3 \wedge d \Csl_3,}
where $M_{12}$ is a twelve-manifold with boundary $M_{11}$. In
the absence of fivebranes we have $G_4 = d C_3$ and
$d G_4 = 0$. In the presence
of fivebranes we have argued above that this equation should
be modified to $d \Gsl_4 = d \rho \wedge e_4/2$. This requires
that we modify the relation between $G_4$ and $C_3$. The modified
Bianchi identity is satisfied with
\eqn\zzd{ \Gsl_4 = d \Csl_3 + A \rho e_4/2 - B d \rho \wedge e_3^{(0)}/2, }
where locally $C_3$ can be viewed as a small fluctuation field about
the fivebrane\foot{The global definition of $C_3$ is given in the following
section.} and $A+B=1$. Since have smoothed out the fivebrane source
we expect on physical grounds that $C_3$ and $G_4$ should be smooth
on the fivebrane, and in fact in the treatment of \wittfive\ it is important
that $C_3$ be well defined on the fivebrane. Since $\rho e_4$ is singular
at the fivebrane this requires that we take $A=0$ and hence $B=1$.

We thus have
\eqn\zze{\Gsl_4 = d \Csl_3 - d \rho \wedge e_3^{(0)}/2 . }
This relation is quite analogous to the relation $H_3 = d B_2 - \omega_3$
which occurs in $D=10$, $N=1$ supergravity coupled to gauge theory and
is central to the Green-Schwarz anomaly cancellation mechanism.
In particular, the relation \zze\ implies that  $C_3$ must have an
anomalous variation under $SO(5)$ gauge transformations in order that
$G_4$ be gauge invariant,
\eqn\zzf{\delta  \Csl_3 = - d \rho \wedge e_2^{(1)}/2. }

Given the modified relation between $G_4$ and $d C_3$
we must ask how the Chern-Simons term should be modified.
These modifications will involve higher derivative
metric interactions. Actually, describing these
terms as higher derivative terms is slightly misleading
since it presupposes a local description of the physics. 
However $\sigma_3$ is not local 
in the metric, since we use
the exponential map to transfer forms from the 
total space of the normal
bundle to a neighborhood of the fivebrane.  
We expect that an eventual
microscopic derivation will explain this nonlocality
or replace it with a local description. 
With this caveat in mind, there
are many higher derivative terms one could add
to the supergravity action. These are constrained by
physical principles such as
supersymmetry  and gauge  invariance.
Here we will only examine the constraints of
gauge invariance under diffeomorphism and 3-form
gauge transformations. We introduce the expression
$\sigma_3$ defined by:
\eqn\zzg{\Gsl_4 - \rho e_4/2 = d( \Csl_3 - \rho e_3^{(0)}/2)
\equiv d(\Csl_3 - \sigma_3). }
(Note that $G_4$ is not exact.)
A natural set of higher order terms relevant to the
anomaly cancellation problem is obtained by replacing
$C_3$ by $\sigma_3$ or $dC_3$ by $G_4, \rho e_4,$
or $d \sigma_3$. One finds in this way twelve
linearly independent  higher derivative
metric interactions. One combination of higher
derivative terms which maintains the Chern-Simons structure
of the original interaction is
\eqn\zzi{S'_{CS} = \lim_{\epsilon \rightarrow 0} - {2\pi \over 6} \int_{M_{11}
-
D_{\epsilon}(W_6)} (\Csl_3 - \sigma_3) \wedge d(\Csl_3 - \sigma_3)
\wedge d(\Csl_3 - \sigma_3 ) }
and we will  take \zzi\ as
the modified Chern-Simons term. It includes higher
derivative interactions involving up to eleven derivatives
of the metric. Moreover, it is not gauge
invariant by itself under diffeomorphisms.
Under diffeomorphisms ($SO(5)$-gauge transformations
of the normal bundle)
the variation of $C_3$ leads to a variation
of \zzi.
Indeed, it follows from \zzf\ that
\eqn\zzh{ \delta (\Csl_3 - \sigma_3) = - d( \rho e_2^{(1)}/2 ) .}
Computing the variation we have
\eqn\zzj{\delta S'_{CS} = \lim_{\epsilon \rightarrow 0}
{2\pi \over 6} \int_{M_{11} - D_{\epsilon}(W_6)} d(\rho e_2^{(1)}/2)
\wedge d(\Csl_3 - \sigma_3)
\wedge d(\Csl_3 - \sigma_3 ). }
%
Integrating by parts and taking the limit and using the fact that
$G_4$ and $C_3$ are smooth near the fivebrane we obtain
\eqn\zzk{\delta S'_{CS} = - {2\pi \over 6} \int_{S_{\epsilon}(W_6)} {e_4 \over
2}
\wedge
{e_4 \over 2} \wedge {e_2^{(1)} \over 2}, }
which by \zzb\ cancels the remaining anomaly in diffeomorphisms of the
normal bundle. In \wittfive\ the cancellation of antisymmetric tensor
gauge transformation of $C_3$ was also studied. It is not hard to see
that the modification we have made to the Chern-Simons coupling preserves
the cancellation found in \wittfive.

\newsec{Global Structure}

Far from the fivebrane $ d \rho $ vanishes and we have locally
$G_4 = d C_3$. On the other hand we have by the definition of a
fivebrane that
\eqn\zzl{\int_{S^4_{\infty}} \Gsl_4 = 1 }
so $C_3$ cannot be globally well defined. Rather, we must define $C_3^i$
in patches of an open cover $\CU_i$ and relate the $C_3^i$ across patches by
antisymmetric tensor gauge transformations, $C_3^i - C_3^j = d \Lambda^{ij} $.
The appropriate machinery for this construction is the \v{C}ech de Rham
complex. A very readable account aimed towards physicists can be found
in \orlando. The situation here is more complicated due to the mixture
of tensor gauge transformations and diffeomorphisms required by the
Green-Schwarz like structure.
Because of the way we have smoothed out the fivebrane source the
quantity with constant linking number through an $S^4$ surrounding
the fivebrane is $\Gsl_4 - \rho e_4/2$. In particular
\eqn\zzm{\int_{S^4_r}  \Gsl_4}
varies from 1 to zero as $r$ decreases from large $r$ to small $r$.  To
explain the consequences for~$C_3$ we first describe the patching conditions
on~$C_3$ near the fivebrane.  Then we summarize briefly a global description;
see \freed\ for a more leisurely exposition.

Choose an open cover $\CV_\alpha$ for the fivebrane~$W_6$. Then in the
transverse space which we take to be $\IR^5$ for simplicity we choose radial
coordinates and split $S^{4}= S^{4}_+ \cup S^{4}_-$ into northern and
southern hemispheres.

We then have patches
\eqn\zzn{\eqalign{ & \CV_\alpha \times S^{4}_+ \times [r>0]  \cr
& \CV_\alpha \times S^{4}_- \times [r>0] \cr }}

The antiderivative $e^{(0)}_{3}(\Theta, \hat y)$
transforms across the overlap region $S^{3} \times I
\cong S^{4}_+ \cap  S^{4}_-$ by an $SO(5)$
gauge transformation $g_{\pm}$.

Using the fact that  $C_{3}$ is well defined on $W_6$ and the
gauge invariance of $G_{4}$ we therefore take:
\eqn\ceepatch{
\Csl^+_{3}  - \Csl^-_{3}
= {1 \over 2} (1+ \rho) d e^{(1)}_{2}(g_\pm, \Theta_+, \hat y_+),
}
where $e_2^{(1)}$ is the integrated form of the cocycle $e_2^{(1)}$
given in \deterb.
Note that this is {\it not} a an antisymmetric tensor gauge transformation.
But we can use the fact that $C_{3}$
has picked up a diffeomorphism variation
to write this as:
\eqn\ceepatchi{
\eqalign{
\Csl^+_{3}  - \Csl^-_{3}
& = d\Biggl[
{(1+ \rho) \over 2}  e^1_{(2)}(g_\pm, \Theta_+, \hat y_+) \Biggr] \cr
& - {1 \over 2} d \rho \wedge e^1_{(2)}(g_\pm, \Theta_+, \hat y_+)\cr}
}
which is a sum of an antisymmetric tensor gauge transformation
and a diffeomorphism gauge
transformation. We thus find that $C_3$ is well defined near
$W_6$. As we move away from the fivebrane $C_3$ requires non-trivial
transition functions between patches which are a combination of
antisymmetric tensor gauge transformations and diffeomorphisms.
At infinity these reduce to  pure antisymmetric tensor gauge
transformations.

A global discussion may be framed in terms of a general ``$\Gamma
$-calculus,'' which is an extension of the usual calculus of differential
forms.  We describe it in terms of an open cover~$\{U_i\}$ of~$M$ which is
{\it good\/} in the sense that all intersections
  \eqn\ocover{U_{i_0\dots i_p} = U_{i_0}\cap\dots \cap U_{i_p}}
are contractible.  The set of good covers is contractible in a suitable
sense, so the particular choice of good cover does not affect the result of
any computation.  Let $\bigl(\Gamma ^\bullet(M),D \bigr)$ be the total
complex of the modified \v Cech-de Rham complex
\def\bigstrut{\vrule height12pt depth3pt width0pt}
\def\smallstrut{\vrule height6pt depth3pt width0pt}
\def\tinystrut{\vrule height1pt depth3pt width0pt}
  \eqn\Gcalc{
 \vtop{\offinterlineskip
      \halign{\hfil#\hfil\bigstrut\quad&\vrule\quad \hfil #\hfil\quad& \hfil
     #\hfil\quad& \hfil #\hfil\quad&\hfil #\hfil\quad& \hfil #\hfil\quad&\hfil
     #\hfil\quad& \hfil #\hfil \cr
      $\vdots$&\cr
      \omit&\smallstrut\cr
      $\Omega ^2(U)$&\cr
      \omit&&&$d\uparrow$\cr
      $\Omega ^1(U)$&&\hbox to 8pt{}&&${\textstyle\to}\atop{\textstyle\delta
        }$\cr
      \omit&\smallstrut\cr
      $C^{\infty}(U,\IT)$&\cr
      \omit&\tinystrut\cr
      \noalign{\hrule}
      &$U_i$&\omit&\omit$U_{ij}$&&$U_{ijk}$&&$\;\;\cdots$\cr
      }}
}
Here $\delta $~is the usual \v Cech differential, the columns form the de Rham
complex modified by replacing 0-forms by circle-valued functions, and
$D = \delta \pm d$ is the total differential.  There is a subspace $\Theta
^p(M)\subset \Gamma ^p(M)$ of ``connection-like'' elements:
  \eqn\Thet{\Theta ^p(M) = \{\omega \in \Gamma ^p(M):D\omega \in \Omega
     ^{p+1}(M)\}.}
For example, an element $\omega =\{g_{ij},\alpha _i\}\in \Theta ^1(M)$
has the form
  \eqn\conncirc{
 \vtop{\offinterlineskip
      \halign{\hfil#\hfil\bigstrut\quad&\vrule\quad \hfil #\hfil\quad& \hfil
     #\hfil\quad& \hfil #\hfil\quad&\hfil #\hfil\quad& \hfil #\hfil\quad&\hfil
     #\hfil\quad& \hfil #\hfil \cr
      2&$\Omega $\cr
      \omit&$\uparrow$\cr
      1&$\alpha _i$&$\to$&0\cr
      \omit&&&$\uparrow$\cr
      0&&&$g_{ij}$&$\to$&0\cr
      \omit&\tinystrut\cr
      \noalign{\hrule}
      &$U_i$&\omit&\omit$U_{ij}$&&$U_{ijk}$\cr
      }}
}
It represents a circle bundle with connection: $g_{ij}$~are
the transition functions of some local trivializations, $\alpha _i$~are the
local connection forms, and $\Omega =D\omega $~is the curvature.
Intuitively, $\omega \in \Theta ^p(M)$~for $p>1$ is a higher degree version
of a connection on a circle bundle.  An
element~$\sigma \in \Gamma ^{p-1}(M)$ is a trivialization of the trivial
bundle $D\sigma \in \Theta ^p(M)$.

Given the fivebrane $W_6\subset M_{11}$ we first define a Poincar\'e dual
form
  \eqn\poindual{\Omega (g)\in \Omega ^5(M)}
by the right hand side of~\nine.
It depends ``functorially'' on the metric~$g$ on~$M$: the construction is
invariant
under diffeomorphisms.  There is also a diffeomorphism-invariant
antiderivative
  \eqn\covder{\mu (g)\in \Omega ^4(M\backslash W)}
on the complement of the fivebrane; it was denoted ``$\rho e_4/2$''
previously.  By working universally we can also choose a
diffeomorphism-invariant connection-like object
  \eqn\hconn{\omega (g)\in \Theta ^4(M)}
which is an antiderivative globally:  it has curvature $D\omega =\Omega $.
Furthermore, there is a trivialization
  \eqn\trivia{\sigma (g)\in \Gamma ^3(M\backslash W)}
off of the fivebrane with ``covariant derivative'' $\mu (g)$: we write
$D\sigma =\mu -\omega $.  The quartet $(\omega ,\sigma ,\Omega ,\mu )$ is our
global description of the smeared-out fivebrane.  It depends on the metric~$g$
and the fixed cutoff function~$\rho $.

In the absence of any fivebranes the 3-form field~$\Csl$ is globally an
element of~$\Theta ^3(M)$ with curvature~$\Gsl_4=D\Csl_3$.  In the presence
of the fivebrane, $\Csl$~is a global trivialization
  \eqn\csl{\Csl_3\in \Gamma ^3(M)}
of~$\omega (g)$ with covariant derivative
  \eqn\gsl{\Gsl_4\in \Omega ^4(M),}
i.e., $D\Csl_3 = \Gsl_4 - \omega (g)$.  With these definitions the modified
Chern-Simons term~\zzi\ makes sense globally and leads to the anomaly
cancellation computed above.

\newsec{Reduction on a Calabi-Yau 3-fold }

One of  the closest relatives of $M$-theory is  $\CN=1$ supergravity in five
dimensions. This theory has
Chern-Simons interactions and chiral strings,
which can lead to
both gauge and gravitational anomalies \fkm. A natural
way to produce such models is via
compactification of $M$-theory on a Calabi-Yau manifold
$X$.
In this case there are $h^{1,1}(X)$ independent
vector fields, including one graviphoton and
$h^{1,1}(X)-1$ vectormultiplets.

Suppose a 5-brane $W$ wraps a four-cycle
$P$ with homology class $[P]=p^A [\Sigma_A] $,
where
$[\Sigma_A]$ is an integral basis for $H_4(X,\IZ)$.
At long distances the  noncompact part of $W$ is
a chiral string in the $\IR^{1,4}$ supergravity.
The number of left and right-moving bosonic  zero modes $N_{L,R}^B$
and right-moving fermionic zero modes $N_R^F$ on the string are given by \msw:
\eqn\strizero{\eqalign{N_L^B &=6D + c_2 \cdot P \cr
N_R^B + {1 \over 2} N_R^F &= 6D + {1 \over 2} c_2 \cdot P \cr
}}
where
\eqn\ctwodotp{c_2 \cdot P = \int_P c_2(TX),}
and the self-intersection is given by
\eqn\inters{D={1 \over 6} \int_X \hat P^3 = D_{ABC} p^A p^B p^C,}
where  $\hat P$ is the Poincar\'e dual class in $H^2(X, Z)$.

The BPS string is magnetically charged under the
gauge field obtained from the Kaluza-Klein reduction
of $C_3$ on $X$ dual to $P$, namely,
$C_3 = C_1 \wedge p^A \theta_A$ where
$C_1\in \Omega^1(\IR^{1,4})$ and
$\theta_A$ is an integral basis of harmonic two-forms on
$X$.
Defining the field  strength associated to
$C_1$ in the presence of a string requires a treatment very similar
to what we presented in the
fivebrane case. In particular, the normal
bundle is now an $SO(3)$ bundle
with connection $\Theta$ and smoothing out the string source requires
that we modify the relation between $G_2$ and $dC_1$ to
\eqn\gslikexfive{
d \Csl_1  = \Gsl_2
+  d \rho \wedge  e_1^{(0)}(\Theta) /2 .
}

The zero mode
spectrum is anomalous and the anomaly is given by descent from
\eqn\strano{{\tilde I_4}(TW, N) = {1 \over 48}\biggl( c_2 \cdot P \left(
p_1(TW) + p_1(N) \biggr)  +  12D p_1(N)  \right).}
The cancellation of anomalies in diffeomorphisms of the tangent
bundle by inflow requires a bulk coupling
\eqn\grinterfive{\Delta S^5 = \int_{M_{5}-D_\epsilon (W)}
C_1 \wedge I_4(TM), }
where now $I_4(TM) =  {1 \over 48} (c_2 \cdot P)  p_1(TM)$.
This coupling can be
obtained by  reduction on $X$ of  the eleven-dimensional bulk term \three.
It
is easy to see that
\eqn\restrfive{\iota^*\biggl(
I_4(TM ) \biggr) = {\tilde I_4}(TW, N)  - {D \over 4} p_1(N),}
where $\iota$ is the inclusion map.
Again there is a part of the anomaly involving gauge transformations
of the normal bundle which is uncancelled
by the inflow.
It is interesting to note that the anomaly of the normal bundle arises only
when  the self-intersection is non-zero which is precisely the
condition for having a
Chern-Simons interaction in five dimensions. Thus  the mechanism for
cancellation of the normal bundle anomaly should be
the same as in $M$ theory.

The modified Chern-Simons coupling is
\eqn\csinterfive{
{S'}^5_{CS} = \lim_{\epsilon \rightarrow 0} -{ 12 D \pi }
\int_{M_{5}-D_\epsilon(W)} (\Csl_1 - \sigma_1) \wedge d( \Csl_1 - \sigma_1)
\wedge d(\Csl_1 - \sigma_1).
}
The anomaly in pure $C_1$ gauge transformations is
compensated by a phase factor coming from
the coupling of $C_1$ to the string worldsheet in a way completely analogous
to the discussion of $C_3$ antisymmetric tensor gauge
transformations in \wittfive.

The cancellation of the
anomaly in gauge transformations of the normal bundle follows
as in the fivebrane case upon application of the relevant version
of the Bott-Catteneo formula:
\eqn\bclower{\int_{S_\epsilon(W_2)} e_2 \wedge e_2 \wedge e_0^{(1)}
= 2 \int_{W_2} (p_1(N))^{(1)} }

It would be interesting to explore whether there are
implications of this discussion for the
black hole entropy following the discussion in \msw.

\bigskip
\centerline{\bf Acknowledgments}\nobreak
\bigskip

We would like to thank  S. Shatashvili for
participation in the beginning of this project
and for many discussions on the fivebrane
anomaly in M-theory. We would also like to
thank R. Bott and E. Witten for essential
discussions and G. Horowitz for discussions
on fivebrane geometry. We also would like to
thank the  Institute for Theoretical Physics
at Santa Barbara
for the stimulating workshop which
made this collaboration possible.
Our joint work there was supported in part by the 
National Science Foundation
under Grant No. PHY94-07194.
The  work  of JH is
supported by NSF Grant No.~PHY 9600697, RM and GM are supported by
DOE grant DE-FG02-92ER40704. DF is supported by NSF grants
DMS-962698, DMS-9304580, and DMS-9627351 and by the
Harmon Duncombe Foundation. 

\appendix{A}{Volume form for all odd rank bundles}

The formulae used in the text for $e_4, e_3^{(0)}$ etc.
have natural extensions to  the  $SO(2n+1)$ case.
The global angular form can be given, as above,
and as in \bottcatt\ as a basic form on $P \times S^{2n}$.
We find:
\eqn\explcten{
\eqalign{
e_{2n}=  {1 \over  2 (4\pi)^n n! }
\sum_{j=0}^{n} (-1)^j {n! \over j!(n-j)!}
\epsilon (F)^j (D\hat y)^{2n-2j}  \hat y, \cr}
}
where
\eqn\shrtfrm{
\epsilon (F)^j (D\hat y)^{2n-2j}  \hat y
\equiv \epsilon_{a_1 \cdots a_{2n+1} } F^{a_1 a_2} \cdots F^{a_{2j-1} a_{2j}}
(D\hat y)^{a_{2j+1}}   \cdots(D \hat y)^{a_{2n}} \hat y^{a_{2n+1}}
}
and the normalization is fixed by noting that
the
volume form defined by $d^{2n+1}y = r^{2n} dr \Omega_{2n}$
has $\int_{S^{2n}} \Omega_{2n} = 2 \pi^{n+1/2}/\Gamma(n+1/2)$.

Similarly, the Chern-Simons form for
the general $SO(2n+1)$ case is given by:
\eqn\moregen{\eqalign
{  e_{2n-1}^0(\Theta_1) &-
e_{2n-1}^0(\Theta_0) \cr
&= - {\epsilon \over  2 (4\pi)^n n! }
  \int_{0}^{1}{dt}
\sum_{j=0}^{n-1} (-1)^j {n! \over j!(n-j-1)!}
\Theta (F_t)^j (D_t \hat y)^{2n-2j-2}  \hat y, \cr}}
where
$\Theta_t = t \Theta_1 + (1-t) \Theta_0 $, and
$F_t = d \Theta_t - \Theta_t^2 $ and $D_t = (d - \Theta_t ).$
The equations simplify
for the case that the normal bundle is topologically trivial.
In that case there is a canonical choice of basepoint
connection $\Theta=0$ for which we may
take an antiderivative of the volume form of
the sphere $S^{2n}$.

It is also worth noting that one can give
$e_{2n}$ a Mathai-Quillen-like representation.
We can introduce  $2n+1$ orthonormal antighost
zeromodes and write, up to a constant, the angular
form for the odd rank case as:
\eqn\angformod{
e_{2n}(g) = {1 \over  2 (2\pi)^n   }
\int \prod_{a=1}^{2n+1} d \rho^a
\exp\Biggl[ \rho^a \rho^b F^{ab} - (D \hat y)^a \rho^a +
\rho^a \hat y^a \Biggr].
}
The MQ representative
of a rapid-decrease Thom class of
odd-rank bundles is therefore:
\eqn\angformodd{
\Phi^{\rm universal}
 =  \kappa_n  e^{-(y,y)} \int \prod_{a=1}^{2n+1}
{d \rho^a \over  \sqrt{2 \pi}}
\exp\Biggl[ \rho^a \rho^b \phi^{ab} - (D  y)^a \rho^a +
\rho^a   y^a \Biggr],
}
where $\kappa_n$ is a normalization constant
and $\phi$ is in the  Weil algebra.
The expression \angformodd\  closely resembles
the Mathai-Quillen representative of the universal
Thom form of even rank bundles, which is a starting
point for the development of topological field theory
(see, for example, \lhl). It would be interesting to
see if the above expressions could also be used
to develop new topological field theories.

\listrefs
\end